\documentclass[aps,prl,reprint,groupedaddress]{revtex4-1}
\usepackage{graphicx,color}
\usepackage{amssymb}   % for math
\usepackage{amsmath}
\usepackage{epstopdf}
\usepackage[hidelinks]{hyperref}
\usepackage{bm}
\usepackage[normalem]{ulem}

\begin{document}
	
	\title{Platform of chiral Majorana edge modes and  its quantum transport phenomena}
	
	\author{James Jun He$ ^1 $, Tian Liang$ ^1 $, Yukio Tanaka$ ^2 $, Naoto Nagaosa$ ^{1,3} $}
	
	\affiliation{
		$ ^1 $ RIKEN Center for Emergent Matter Science (CEMS), Wako, Saitama 351-0198, Japan\\	
		$ ^2 $Department of Applied Physics, Nagoya University, Nagoya, Japan\\
		$ ^3 $Department of Applied Physics, The University of Tokyo, Tokyo 113-8656, Japan
}
	
	\date{\today}
	
	\begin{abstract}

    We propose a method to create two-dimensional topological superconductors with a heterostructure of ferromagnet (FM), topological insulator (TI) thin film and superconductor, in which the two surfaces of the TI thin film are treated as a two-dimensional system. One of surfaces is superconducting due to proximity effect and the other feels an exchange field from the FM. We show that there is a topological phase with single chiral Majorana edge mode that exists in readily achievable parameter regions and does not require magnetization to be small. An experimental setup is proposed based on our model to uniquely determine the existence of Majorana chiral modes using a Josephson junction. Also, we show that multiple chiral Majorana edge modes may appear when unconventional superconductors are used.

	\end{abstract}

	\maketitle
	
	\paragraph{Introduction ---} Majorana fermions in condensed matter systems \cite{Wilczek, Kitaev2001}, or Majorana quasi-particles, have been the subject of intense research due to their exotic properties such as fractional Josephson effect \cite{Kitaev2001,Tanaka2,Tanaka3,Kwon, Fu1, Lutchyn, Vic2011}, resonant Andreev reflection \cite{Vic2009,Wimmer,Tanaka1,Bolech}, enhanced cross Andreev reflection \cite{JamesNC}, spin selection \cite{Simon, JamesPRL, Oreg,XLiu,Lunhui, Noah},  non-Abelian statistics \cite{RG, Ivanov, Fujimoto, STF, Alicea2011},  etc. Tremendous theoretical \cite{Fu2008,Sau2010,Lutchyn,Alicea2010, ORV, Potter,TP,Sho,Vic2018,Ruixing} and experimental \cite{Mourik,Deng,Das,Yazdani2014,JiaPRL, Marcus1,Marcus2, Yazdani2017, Kouwenhoven2018} progress has been achieved to create zero-dimensional Majornana bound states in one-dimensional topological superconductors (1D TSCs). Promising braiding methods have been proposed based on such 1D systems \cite{AliceaNP} for the final proof of the non-Abelian particles, as well as for the creation of topological quantum computers \cite{Kitaev2003,Nayak}.		
		
	On the other hand, 1D propagating Majorana modes at the edges of 2D TSCs have been realized recently  in an experiment \cite{Qinglin2017} that combines quantum anomalous Hall insulators (QAHIs) with superconductors (SCs) \cite{Qi2010}. In such a system, the TSC with single chiral Majorana edge mode appears  only when the out-of-plane magnetization, $ M_z $, is small compared to the superconducting gap $ \Delta $. This parameter regime may be achieved by tuning $ M_z $ with external magnetic field. In this way, narrow regions of half-quantized conductance were observed, which are considered as a signature of the chiral Majorana mode \cite{Chung2011,JWang2015,CZ2017}. 
	
	However, controversy arises based on an observation that half-quantized conductance may also appear trivially without any Majorana mode \cite{Wen2018,Sau2018}  if the edge states of the QAHI go through a long-enough path in the SC so that the SC part behaves just like a metal connecting two QAHIs in series. Due to the requirement of very small magnetization and the fact that the QAHI was obtained by magnetic doping (which induces disorder), domains are likely to form and long paths for the QAHI edge states can exist. This difficulty of distinguishing the TSC explanation from the trivial one stems from the theoretical model where both surfaces of the TI have the same magnetization that competes directly against the superconductivity order parameter, resulting in the requirement $ M_z < \Delta $ \cite{Qi2010}. Thus, Majorana systems beyond this limitation are desired. 	
	
\begin{figure}
		\includegraphics[width=3in]{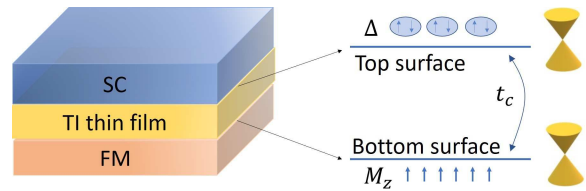}
		\caption{Schematic FM/TI/SC heterostructure. The two surfaces of the TI thin film have Dirac dispersions. One of them is superconducting due to proximity effect and the other feels an out-of-plane exchange field from the FM.}
		\label{Hetero}
	\end{figure}

	In this letter, we propose an alternative method to create 2D TSCs with a heterostructure of ferromagnet (FM), topological insulator (TI) thin film and superconductor (SC), in which the two surfaces of the TI thin film form a two-dimensional system (FIG.\ref{Hetero}). One surface is superconducting due to proximity effect and the other feels an exchange field from the FM. We show that there is a topological phase with single chiral Majorana edge mode that exists in readily achievable parameter regions and does not require magnetization to be smaller than the SC gap. An experimental setup containing a Josephson junction is proposed  to uniquely determine the existence of Majorana chiral modes, in which a smoking-gun evidence is a change of conductance from $ 1/2 \leqslant \sigma_{12} \leqslant 1 $ to $ \sigma_{12}=1/3 $ (the unit of conductance is $ e^2/h $ throughout this manuscript) as the current is tuned up across the Josephson critical current. We also show that multiple chiral Majorana edge modes may appear when unconventional superconductors are used.
	
	\paragraph{Model --- }
	Assuming the FM to be insulating, the low-energy properties of the system is determined by the two surface states of the TI thin film. The two surfaces, however, experience different environments. The bottom surface is in good contact with the FM and  feels a strong out-of-plane exchange field, whereas the top surface becomes superconducting due to proximity of the SC, as shown in FIG.\ref{Hetero}. In the Nambu basis $ \{ c_{t,\uparrow}(\bm k),c_{t,\downarrow}(\bm k),$  $c_{b,\uparrow}(\bm k),c_{b,\downarrow}(\bm k), $  $c_{t,\uparrow}^\dagger(-\bm k),c_{t,\downarrow}^\dagger(-\bm k),$  $c_{b,\uparrow}^\dagger(-\bm k),c_{b,\downarrow}^\dagger(-\bm k)\} $, the effective Hamiltonian of the system is 
	%$ \hat{H} = \sum_{\bm k} H_{ij}(\bm k) c_i^\dagger(\bm k) c_j(\bm k) $ with
	\begin{align}
	H(\bm k) = & v(k_x s_y \tau _z-k_y s_x \tau _0 ) \sigma_z + M_z s_z \tau _z \frac{\sigma_z - \sigma_0 }{2} \notag \\ 
	 +& t_c s_0 \sigma_x \tau_z - s_0 (\mu  \sigma_0+\delta E \sigma_z) \tau_z \notag\\
	+ & [\Delta_{s}(\bm k) + \bm d(\bm k) \cdot \bm \sigma]i s_y \tau_y \frac{\sigma_z+\sigma_0}{2} .  \label{EqH}
	\end{align}
	The Pauli matrices $ s_{x,y,z} ,  \sigma_{x,y,z},  \tau_{x,y,z}$ act on spin, layer and particle-hole spaces respectively. $ M_z $ is the exchange field felt by the bottom layer, $ \Delta(\bm k) $ and $ \bm d(\bm k) $ are the singlet and triplet SC order parameters on the top layer. The constant $ t_c $ is the hybridization energy between the two layers, which depends on the film thickness. $ \mu $ is the chemical potential while $ \delta E $ denotes the energy shift between two surfaces.
	We assume the decay length of the exchange field to be smaller than the TI film thickness so that the top surface does not couple with the exchange field directly, although an indirect coupling can be conveyed by the hybridization $ t_c $. 
		
	The only discrete symmetry of this system is the redundant particle-hole symmetry and thus it belongs to D class \cite{Schnyder,Teo} in which topological phases in two dimensions can be identified by Chern numbers. This model describes a currently accessible experimental system where the heterostructure can be fabricated by molecular beam epitaxy (MBE) method  with controllable thickness \cite{Mogi}. 	Note that it becomes a QAHI if the SC on the top surface is replaced by another FM same  as the bottom surface \cite{Yu}. This property is useful when we discuss junctions of QAHIs and TSCs. 
	
	\begin{figure}
		\includegraphics[width=3.3in]{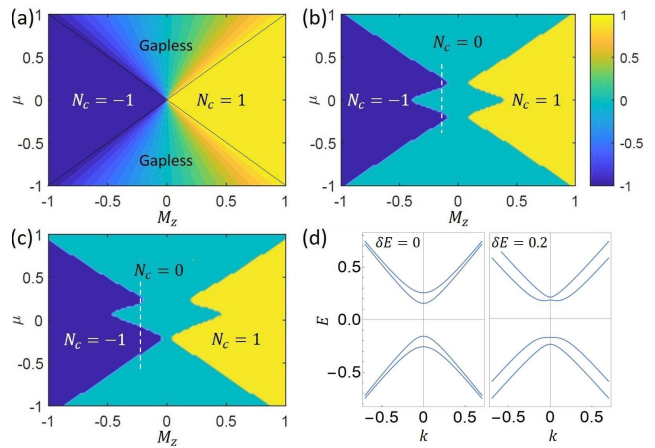}
		\caption{ %Three gapped phases exist with Chern number $ N_c=1  $ (yellow), $ N_c=-1 $ (dark blue) and $ N_c=0 $ (green) respectively.
		Phase diagrams for the FM/TI/SC heterostructure. The color denotes the summation of Berry curvature of all the occupied electronic states divided by $ 2\pi $. It is an integer (Chern number $ N_c $) when the system is gapped. $v=1, \Delta_s=0.1 $ for all the calculations. (a) $ \delta E=0, t_c=0 $. There are parameter regions where the system is gapless and the colors there between dark blue and yellow indicate the crossover between the two gapped regions schematically. The black lines are boundaries between the gapped and gapless phases. (b) $ \delta E=0, t_c=0.2 $. (c) $ \delta E =0.2, t_c=0.2$. (d) Schematic normal state band structures with $ \delta E=0 $ (left) and $ \delta E=0.2 $ (right). They resemble Rashba bands when $ \delta E $ is large.}
		\label{sWave}
	\end{figure}
	
	\paragraph{With s-wave SCs --- }
	%Let's  first consider the simplest (and the most experimentally accessible) situation when the SC is s-wave. 	
	In the limit $ t_c = 0$, the stacked TI surfaces may be unfolded and regarded as two sections of one surface aligned side by side.  The edge of the stacked system then becomes the boundary between the sections. When they are gapped by superconductivity and magnetization respectively, a chiral Majorana mode appears at the boundary \cite{Fu2009}. Similarly, such a  mode is expected at the edge of the heterostructure in FIG.\ref{Hetero}. The phase diagram obtained by calculating the total Berry curvature $ \gamma $ is shown in FIG.\ref{sWave}(a). In normal states, $ \gamma $ corresponds to  Hall conductance. If a  bulk gap exists, it is quantized so that $ N_c = \gamma/2\pi $ with the integer $ N_c $ being the Chern number \cite{TKNN}. In superconducting states, the Chern number can be defined in the same mathematical way using the Bogoliubov-de Gennes Hamiltonian  (although it is no longer related to Hall conductance). 
	For $ |\mu| < |M_z| $, the system is gapped and $ N_c=\text{sign}[M_z] $,  corresponding to a single chiral Majorana edge mode whose chirality is determined by the  magnetization direction. 
	If $ |\mu| > |M_z| $, the bottom surface becomes gapless and the total Berry curvature is not quantized. 
	
	When $ t_c \neq 0 $, it competes with $ M_z $ trying to generate a trivial hybridization gap. Consequently, larger $ |M_z| $ is required to obtain a non-zero Chern number at $ \mu=0 $, as shown in FIG.\ref{sWave}(b). If $ M_z>0 $ is small and $ \mu $ deviates from zero, the system first enters a non-trivial region with $ N_c=\text{sign}[M_z] $ and then transitions into a trivial phase again for large $ |\mu| $, as shown by the dashed line. This is easily understood by looking at the normal state band structures as shown in the left panel of FIG.\ref{sWave}(d). The bands are non-degenerate due to the inter-layer hybridization $ t_c $ and the exchange field $ M_z $. The  two subbands closest to zero energy have a trivial gap opened by $ t_c $ and thus SC has no effect if $ \mu=0 $, giving $ N_c=0 $. As $ |\mu| $ increases, the Fermi level cuts a single band and the resulting SC is topological with single chiral Majorana edge mode. As $ |\mu| $ further increases, the Fermi level cuts two bands and the system becomes a trivial SC again. 
	
	This even-odd effect of the number of Fermi surfaces resembles that of Rashba systems \cite{Lutchyn,Potter}. The similarity becomes clearer in band structure when a difference in chemical potentials of the top and bottom surfaces, $ \delta E = 0.2 $, is considered. As shown in the right panel of FIG.\ref{sWave}(d), the conduction (valence) band looks just like a Rashba band with positive (negative) mass and a Zeeman field which splits the degeneracy at $ k=0 $. However, the splitting of the valence band is larger than that of the conduction band. This is because the states of the conduction (valence) band near $ k=0 $ are from the top (bottom) surface of the TI film, and the top surface only couples with the FM indirectly through $ t_c $ while the bottom surfaces feels the exchange field directly. Consequently, the topological region (where the number of Fermi surfaces is odd) is wider when $ \mu<0 $, as shown by the dashed line in FIG.\ref{sWave}(c) where the whole phase diagram is obtained.		
	The effect of $ \delta E $ has been discussed \cite{Morimoto} and it  happens in real experiments \cite{Yoshimi}. Note that no new phases emerge from this energy difference of two surfaces and its effect on the phase diagram is only quantitative.

\begin{figure}
		\includegraphics[width=3.1in]{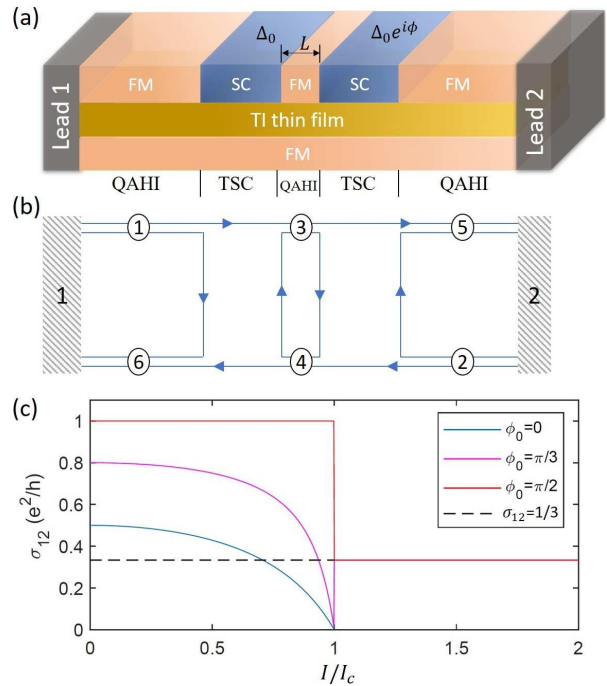}
		\caption{ (a) Experimental setup including a Josephson junction of two TSCs. (b) The Majorana edge modes corresponding to the system in (a). (c) Two-terminal conductance $ \sigma_{12} $ as functions of the applied current $ I $. $ I_c $ is the Josephson critical current and $ \phi_0 = k_F L $ is the kinetic phase acquired by propagating across the junction. Note that all the curves coincide when $ I/I_c>1 $.
		}
		\label{junction}
	\end{figure}

	\paragraph{Experimental detection ---}
	%Since Majorana modes exist as in-gap states of superconductors, detection by charge transport experiments usually requires a weak link between the sample and the electrode so that the contribution of the bulk Cooper pairs can be neglected. Such weak links can be achieved by several methods such as scanning tunneling microscopy (STM), point contact, gating, etc. Alternatively, instead of connecting the lead and the sample directly, one may make an intermediate region where there are very few or even single electronic channel. In two  dimensions, quantum spin Hall insulators  and quantum anomalous Hall insulators (QAHIs) are good candidates for such few-channel region.  Due to the time reversal symmetry breaking in our model, it is natural to use QAHIs. However, instead of a single QAHI-TSC-QAHI junction  \cite{Chung2011,JWang2015,Qinglin2017}, we propose an experimental setup including a Josephson junction \cite{CZ2018,Shen2018}, as shown in FIG.\ref{junction}(a). As we shall see, a simple conductance measurement with  this junction can distinguish the topological phase of the superconducting region from other trivial explanations and thus provide smoking-gun evidence for the chiral Majorana edge modes. 
	The chiral Majorana modes in the heterostructure discussed above may be detected by a QAHI-TSC-QAHI junction as shown by previous researches \cite{Chung2011,JWang2015,Qinglin2017}. To avoid trivial explanations \cite{Wen2018,Sau2018} and uniquely determine the existence of chiral Majorana modes, however, we propose an experimental setup including a Josephson junction \cite{CZ2018,Shen2018}. As shown in FIG.\ref{junction}(a), the setup is achieved by adding SCs and FMs alternately on the top surface of the TI thin film while attaching a uniform FM to the bottom surface. Regions with both top and bottom surfaces coupled to FMs are QAHIs \cite{Yu} while those with one surface coupled to FM and the other to SC are TSCs with single Majorana edge mode, as we have discussed. The two TSCs form a Josephson junction, which are connected to external electrodes 1 or 2 through another QAHI at each end. Note that the SCs are \emph{not grounded}, contrary to that of reference \cite{Shen2018}. Figure \ref{junction}(b) schematically shows how the edge states propagate in Majorana basis in which each normal edge state of the QAHIs is regarded as two Majorana states. 	
	As we shall see in the following, a simple measurement of the current-dependent conductance from lead 1 to lead 2  provides smoking-gun evidence for the chiral Majorana edge modes.

	Consider a current  $ I < I_c^{bulk} $ ($ I_c^{bulk} $ being the critical current of the bulk SCs) flowing from lead 1 to lead 2. When $ I > I_c $,  where $ I_c $ is the Josephson critical current, the current through the junction is normal and a voltage difference between two SCs exists. The current across the junction is carried by the normal edge states of the QAHI and thus it is determined by the occupation numbers of the edge states. We can define the chemical potential $ \mu_i $ ($ i=1,2,...,6 $) for each QAHI edge state as shown in FIG.\ref{junction}(b). They must satisfy	$\mu_1-\mu_6 = \mu_3-\mu_4=\mu_5-\mu_2= e I $ due to the quantized Hall conductance of the QAHIs. Additionally, the left half of the system is a QAHI-TSC-QAHI junction whose scattering matrix has been obtained \cite{Chung2011}. A simple application of the scattering coefficients to multi-terminal measurement leads to $\mu_3=\mu_6=(\mu_1+\mu_4)/2$. Similarly, we get $\mu_4=\mu_5=(\mu_2+\mu_3)/2$ for the right half of the system.  	
	Combination of these relations leads to
	\begin{align}
    \sigma_{12}^> = \frac{e I} {\mu_1-\mu_2}= \frac{1}{3}, \text{ when } I > I_c.
	\end{align}
	This result turns out to be the same as the case where the SCs are replaced by normal metals. However, it should be emphasized that the current $ I $ we consider here is smaller than the bulk critical current of the SCs and the TSCs with chiral Majorana edge states remain intact.

	When $ I < I_c $, the current flowing through the junction (the middle QAHI region) is a supercurrent carried by Cooper pairs and the aforementioned relations of $ \mu_i $ no longer hold. There is no voltage drop between the two TSCs and they can only differ by a phase of the order parameter $ \phi $ which is related to the current by $I=I_c \sin (\phi/2) $ or $ \phi=2 \arcsin (I/I_c) $. The existence of $ \phi $ can drastically affect $ \sigma_{12} $ by inducing interference between different Majorana paths and thus changing the tunneling amplitude \cite{Shen2018}. Taking this into account, one obtains the conductance
	\begin{align}
     \sigma_{12}^< 
     %=  (\cos^2 \phi_0+\cos^2 \frac{\phi}{2})^{-1} \cos^2 \frac{\phi}{2} 
     =  \frac{1-(I/I_c)^2}{2-(I/I_c)^2-\sin^2 \phi_0}, \text{ when } I < I_c, 
	\end{align}	
	where $ \phi_0 = k_F L $ is the kinetic phase acquired by the edge states across the junction \cite{Shen2018}.  ($ k_F $ is the Fermi wave vector and $ L $ is the length of the junction.) 
	When $ I\ll I_c $, $ \phi $ is small and its effect on $ \sigma_{12} $ is negligible. As $ I $ approaches $ I_c $, $ \phi $ increases from zero to $ \pi $ and $ \sigma_{12} $ decreases to zero, as shown in FIG.\ref{junction}(c). Note that in the special case with $ \phi_0=\pi/2 + n\pi $, $ \sigma_{12}^< $ is constantly unity. 
	
	In summary, as the current increases, the two-terminal conductance $ \sigma_{12} $ starts with a value between $ 1/2 $ and $ 1 $ and decreases until it exceeds the Josephson critical current, above which $ \sigma_{12} $ is constantly $ 1/3 $.	This is a unique consequence of the chiral Majorana edge states in the TSCs. 

%\begin{figure}
%	\includegraphics[width=3.6in]{FIG3c.eps}
%\end{figure}

	\paragraph{Multiple Majorana modes --- } 
	With the setup in FIG.\ref{Hetero}, it is interesting to consider unconventional  SCs rather than s-wave \cite{Linder}. Particularly, previous studies show that multiple chiral Majorana edge modes may be achieved in two-dimensional Rashba system using p-wave and d-wave SCs with Zeeman field \cite{Yanase1,Yanase2,Yada}. %In the following, we explore the similar idea in our model in Eq.(\ref{EqH}).
		
	Consider a general pairing potential $ \hat{\Delta}(\bm k) = (\Psi +  \hat{\bm d} \cdot \bm \sigma ) i \sigma_y $ that acts on TI surface states with the Hamiltonian $ \hat{h}(\bm k) =  \bm g \cdot \bm \sigma + V_z \sigma_z - \mu_0 $ where $ V_z $ is a perpendicular Zeeman field and $ \bm g = (k_y, -k_x,0)$ is the spin-orbit coupling vector.
	Let us, for clarity, assume the Fermi level to be above the band crossing point ($ \mu_0 >0 $), then the superconductivity gap function in the band basis becomes $ \Delta_+ = e^{i\theta}[(d_x \cos \theta + d_y \sin \theta)\sin \alpha + i (d_y \cos \theta - d_x \sin \theta- \Psi \cos \alpha) ]   $ with $ \alpha=\arcsin (V_z/\sqrt{k^2+V_z^2}) $ and $\theta = \arg (k_x + i k_y) $. 
	
	\begin{figure}
		\includegraphics[width=3.2in]{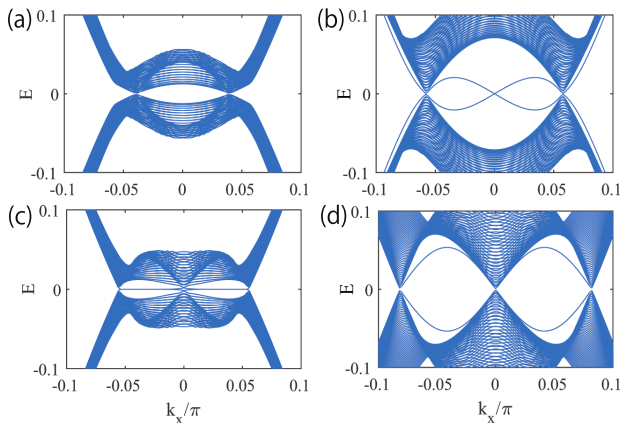}
		\caption{ The energy dispersion with terminated	edges in y-direction for d-wave pairing. Only the lowest 100 energy levels are shown. The y-direction is defined as the [010] direction for (a) and (b) but as the [110] direction  for (c) and (d). In (a) and (c), $ M_z=0 $. In (b) and (d), $ M_z=0.4 $. Other parameters are, $ v=1, t_c=0.2, \mu=0.3, \Delta_d=1.9$.}
		\label{FigEdge}
	\end{figure}
	
	If only singlet (s-wave and d-wave) pairings are considered, the gap function on the top surface state is simply $ \Psi \cos \alpha =(\Delta_s + \Delta_d \cos 2\theta)\cos \alpha $. When $ |\Delta_s| > |\Delta_d| $, it is fully gapped and the topological property is the same as the pure s-wave case discussed previously. When $ |\Delta_s| < |\Delta_d| $, it becomes nodal and there may be edge states depending on the direction of the open boundaries, similar to  well-known usual d-wave superconductors \cite{Hu,Ryu,Tanaka2,Tanaka5,Sato,FaWang,Noah2}. 
	For the heterostructure in FIG.\ref{Hetero}, the energy dispersion on the [010] and [110] edges are shown in FIG.\ref{FigEdge} for pure d-wave. (The s-wave component only shifts the positions of the nodal points.) If $ M_z=0 $, this system behaves just as usual d-wave SCs in which flat bands appear in [110] direction (FIG.\ref{FigEdge} (c)) but not in [010] direction (FIG.\ref{FigEdge} (a)). However, when $ M_z $ is turned on, the change of the state is quite different from that of usual d-wave SCs. Particularly, for large $ M_z $, dispersive edge modes appear on [010] edges, as shown in FIG.\ref{FigEdge}(b).

\begin{figure}
		\includegraphics[width=2.6in]{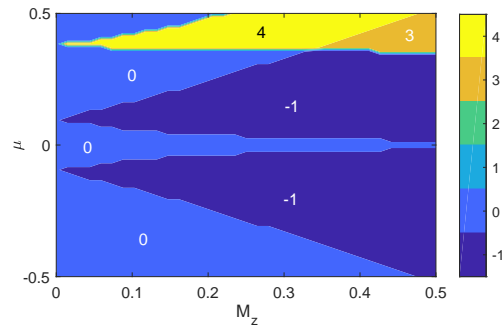}
		\caption{Phase diagram with $\Delta_s=0, \Delta_p=0.1,\bm d =\bm d'= (\sin \theta, \cos \theta,0) $ and $\Delta_d=0.3 $. $ t=1,t_c=0.1$. Colors denotes the Chern numbers.
		}
		\label{dWave}
	\end{figure}

   When p-wave pairing is included, we consider three typical cases, $ \bm d_\parallel = \Delta_p (\sin \theta, -\cos \theta, 0) $, $ \bm d_\perp = \Delta_p (\cos \theta,\sin \theta, 0) $ and $ \bm d'=(\sin \theta, \cos \theta, 0) $. In the first two cases, the d-vectors are either parallel or perpendicular to the spin-orbit vector $ \bm g $. For $ \bm d_\parallel $, the gap function is $ (\Delta_p + \Psi \cos \alpha) $ so that p-wave behaves the same as s-wave. For $ \bm d_\perp $, the gap function becomes $ \Delta_p \sin \alpha - i \Psi \cos \alpha$, which is always fully gapped when  $ \Delta_p V_z \neq 0 $. In the heterostructure, it can support single chiral Majorana mode. With the third choice, $ \bm d' $,  the gap function is $ \Delta_p \sin 2\theta \sin \alpha + i (\Delta_p-\Delta_d \cos \alpha) \cos 2\theta $. Note that, when $ \Delta_p V_z \neq 0 $, this is also fully gapped  except at $ \Delta_p = \Delta_d \cos \alpha $ where the gap changes sign. A phase diagram with this pairing potential ($ \Delta_p=0.3, \Delta_p=0.1 $) is obtained in FIG.\ref{dWave}, where each phase is identified by a Chern number that is equal to the number of chiral Majorana modes. A maximum of four Majorana modes can exist in this system.   
	
	\paragraph{Discussion --- }
	We have proposed a platform of Majorana edge channels by using superconductor/topological insulator/ferromagnet (SC/TI/FM) heterostructures. The topological phase is much easier to be realized compared with the setups studied thus far. The phase diagrams are revealed for s-wave, p-wave, and d-wave pairings for the SC. A smoking gun experiment is also proposed to confirm the Majorana edge channels which exclude the other possibilities. 
	The heterostructures including TI have been already realized experimentally \cite{Mogi,Yoshimi}. By this technique, the quantized anomalous Hall effect is realized at higher temperature due to the suppressed inhomogeneity of the exchange gap \cite{Mogi}. Also the different energy position of the Weyl point between the top and bottom surfaces enables the insulating phase which can support the topological magnetoelectric effect \cite{Yoshimi}. With these artificial structures, one can design various Majorana edge channels to realize the circuits with dissipationless current and even quantum computation \cite{SCZhangPNAS}.  
	
	\begin{acknowledgments}		
	\paragraph{Acknowledgment --- } 
	J.J.H. is very grateful to Chao-Xing Liu for discussions. 
	N.N. was supported by Ministry of Education, Culture, Sports, Science, and Technology
	Nos. JP24224009 and JP26103006, the Impulsing Paradigm
	Change through Disruptive Technologies Program of Council
	for Science, Technology and Innovation (Cabinet Office,
	Government of Japan), and Core Research for Evolutionary
	Science and Technology (CREST) No. JPMJCR16F1 and No. JPMJCR1874, Japan.	
	Y.T. was supported by Grant-in-Aid for Scientic Research
	on Innovative Areas, Topological Material Science (Grants No. No. JP15H05851, No. JP15H05853, and	No. JP15K21717) and Grant-in-Aid for Scientic Research B (Grant No. JP18H01176) from the Ministry
	of Education, Culture, Sports, Science, and Technology, Japan (MEXT).		
	\end{acknowledgments}
	
	\bibliographystyle{apsrev4-1}

\onecolumngrid
\pagebreak

\begin{center}
	\textbf{\large Supplementary }
\end{center}

%%%%%%%%%% Prefix a "S" to all equations, figures, tables and reset the counter %%%%%%%%%%
\setcounter{equation}{0}
\setcounter{figure}{0}
\setcounter{table}{0}
\setcounter{page}{1}
\makeatletter
\renewcommand{\theequation}{S\arabic{equation}}
\renewcommand{\thefigure}{S\arabic{figure}}
\renewcommand{\bibnumfmt}[1]{[S#1]}
\renewcommand{\citenumfont}[1]{S#1}
%%%%%%%%%% Prefix a "S" to all equations, figures, tables and reset the counter %%%%%%%%%%

\subsection{\bf Berry curvature and Chern number \\ }
In two-dimensional systems, the total berry curvature of the occupied electron states determines the topological property of the system. It is obtained as \cite{SupBernevigBook}
\begin{align}
	\gamma = -\text{Im} \sum_{E_n<0} \sum_{n'\neq n} \int d^2 \bm k \frac{\langle n |(\partial_{k_x} H_{\bm k}) |n'\rangle \langle n' |(\partial_{k_y} H_{\bm k})|n\rangle - \langle n |(\partial_{k_y} H_{\bm k}) |n'\rangle \langle n' |(\partial_{k_x} H_{\bm k})|n\rangle}{(E_n-E_{n'})^2}.
\end{align}
When the system is gapped, we can define the Chern number as 
\begin{align}
	N_c = \gamma/2\pi.  
\end{align}

\subsection{\bf Derivation of the conductance when $ I > I_c $\\}
Consider the left half of FIG.3(b), which is a QAHI-TSC-QAHI junction. The electron-electron and electron-hole tunneling probabilities from edge channel $ j $ to edge channel $ i $ are \cite{SupChung2011} 
\begin{align}
	T_{31}^{ee} = T_{31}^{he} = T_{61}^{ee} = T_{61}^{he} = 1/4,\\
	T_{34}^{ee} = T_{34}^{he} = T_{64}^{ee} = T_{64}^{he} = 1/4,\\
	T_{16}^{ee} = T_{43}^{ee} = 1.
\end{align}
Other terms vanish. The current-voltage relation (at zero temperature) is given by \cite{SupDatta}
\begin{eqnarray}
	I_i = \sum_{j} K_{ij} (\mu_j - \mu_s)/e,
\end{eqnarray}
with 
\begin{align}
	K_{ij} \equiv \delta_{ij}-T_{ij}^{ee}+T_{ij}^{he},\\
	K_{ii} = 1, K_{16} = K_{43} = -1
\end{align}
and other terms vanish. 
So we have
\begin{align}
	e\left(\begin{array}{cccc}
		I_1\\I_6\\I_3\\I_4
	\end{array}\right) = \left(\begin{array}{cccc}
		1 & -1 & & \\
		0 & 1 & &\\
		& & 1 & 0 \\
		& & -1 & 1 
	\end{array}\right) \left(\begin{array}{cccc}
		\mu_1-\mu_s\\ \mu_6-\mu_s\\\mu_3-\mu_s\\\mu_4-\mu_s
	\end{array}\right) 
	\rightarrow  
	\left(\begin{array}{cccc}
		\mu_1-\mu_s\\\mu_6-\mu_s\\\mu_3-\mu_s\\\mu_4-\mu_s
	\end{array}\right) 
	= e\left(\begin{array}{cccc}
		1 & 1 & & \\
		0 & 1 & &\\
		& & 1 & 0 \\
		& & 1 & 1 
	\end{array}\right)\left(\begin{array}{cccc}
		I_1\\I_6\\I_3\\I_4
	\end{array}\right) ,
\end{align}
Setting $ I_1 = -I_4 = I $ and $ I_3=I_6=0 $, we obtain 
\begin{align}
	\mu_1-\mu_s = -(\mu_4-\mu_s) = e I, \\
	\mu_6-\mu_s =\mu_3-\mu_s =  0, \\
\end{align}
so that
\begin{align}
	\mu_3=\mu_6 = \mu_s = (\mu_1+\mu_4)/2, \\
	\mu_1-\mu_6 = \mu_3 - \mu_4 =  e I.
\end{align}
Similar analysis for the right QAHI-TSC-QAHI junction gives 
\begin{align}
	\mu_4=\mu_5 = \mu'_s = (\mu_2+\mu_3)/2, \\
	\mu_5-\mu_2 = \mu_3 - \mu_4 =  e I.
\end{align}

\subsection{\bf TI surface with general pairing\\}
Consider a general pairing potential $ \hat{\Delta}(\bm k) = (\Psi +  \hat{\bm d} \cdot \bm \sigma ) i \sigma_y $ that acts on a single Dirac cone $ \hat{h}(\bm k) =  \bm g \cdot \bm \sigma + V_z \sigma_z - \mu_0 $. The constant $ V_z $ is a Zeeman field along z-direction and $ \bm g = (k_y, -k_x,0)$ is the spin-orbit coupling vector.
In the band basis, the pairing potential $ \hat{\Delta}(\bm k) $ is transformed to
\begin{align}
	\tilde{\Delta}(\bm k) = (\tilde{\Psi} +  \tilde{\bm {d}} \cdot \bm \sigma ) i \sigma_y
\end{align}
with 
\begin{align}
	&\tilde{\Psi}= -i d_z e^{i \theta }, \label{EqDelta1} \\
	& \tilde{d}_x=  [-\cos \alpha (d_x \cos \theta +d_y \sin \theta)+i \Psi \sin \alpha]e^{i \theta },\\
	& \tilde{d}_y=(d_y \cos \theta -d_x \sin \theta)e^{i \theta } ,\\
	& \tilde{d}_z=[\sin \alpha (d_x \cos \theta +d_y \sin \theta)+i \Psi \cos \alpha] e^{i \theta },\label{EqDelta2}
\end{align}
where we have defined two angles $ \alpha $ and $ \theta $ as 
\begin{align}
	\alpha=\arccos \frac{V_z}{\sqrt{k^2+V_z^2}}, \theta = \arg (k_x + i k_y).
\end{align}
When the Fermi level is not close to the band crossing point, the effect of inter-band pairing ($ \tilde{\Psi} $ and $ \tilde{d}_z $) can be ignored. The remaining intra-band pairing is
\begin{align}
	\Delta_\pm =  i \tilde{d}_y \mp \tilde{d}_x
	= e^{i\theta}[\pm (d_x \cos \theta + d_y \sin \theta)\cos \alpha 
	+ i (d_y \cos \theta - d_x \sin \theta \mp \Psi \sin \alpha) ] .
\end{align} 
Assuming the Fermi level to be above the band crossing point, then the gap function is just $ \Delta_+ $. 

\begin{figure}[h]
	\includegraphics[width=7.2in]{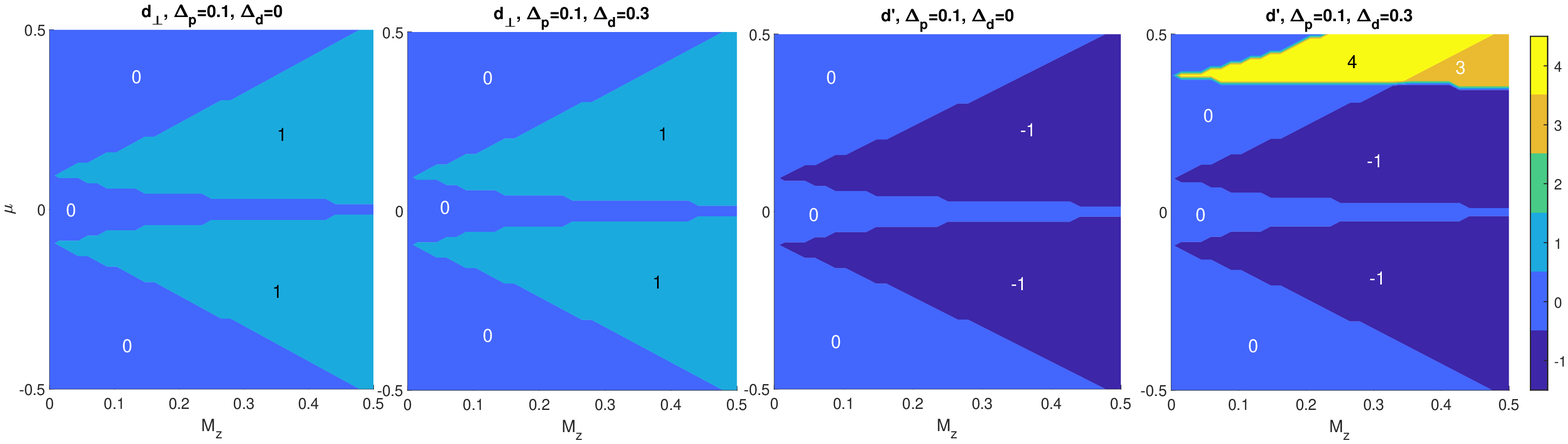}
	\caption{Phase diagrams for different d-vectors ($ \bm d_\perp $ and $ \bm d' $ as defined in the main text) of spin-triplet pairing with and without d-wave component. }
	\label{FIGS1}
\end{figure}

\subsection{\bf More about the phase diagram with unconventional SC\\ }
As comparison, we obtained the phase diagrams for both $ \bm d_\perp $ and $ \bm d' $, as defined in the main text, with and without d-wave component as shown in FIG.\ref{FIGS1}. In the case of $ \bm d_\perp $, the phase diagram is not essentially different from the s-wave case and adding a d-wave component does not change the topological property. For $ \bm d' $, the first observation is the inverted sign of the Chern number, which is due to the particular choice of $ \bm d' $ and the basic topological property is not different from $ \bm d_\perp $. However, adding a d-wave component to $ \bm d' $ triplet pairing changes the topology dramatically and four Majorana modes can appear. This change is expected since increasing d-wave component will reverse the sign of the gap, as discussed in the main text.

%\subsection{Calculation of the edge states --- }We construct tight-binding models corresponding the continuum model in Eq.(1). 

%\clearpage

\end{document}